\documentclass[aps,amsfonts,amsmath,prd,preprint,article,nofootinbib]{revtex4}

\usepackage[toc,page]{appendix}
\newcommand{\beq}{\begin{equation}}
\newcommand{\eeq}{\end{equation}}
\newcommand{\p}{\partial}
\usepackage{amsfonts,amsmath,amssymb,epsfig}
\usepackage{subcaption}
\usepackage{hyperref} 
\hypersetup{ 
	colorlinks=true, 
	linkcolor=black, 
	filecolor=black, 
	citecolor = blue,       
	urlcolor=blue, 
} 
\usepackage{csquotes}
 
\numberwithin{equation}{section}

\newcommand{\nocontentsline}[3]{}
\newcommand{\tocless}[2]{\bgroup\let\addcontentsline=\nocontentsline#1{#2}\egroup}

\begin{document}
	\title{The tunneling wavefunction in Kantowski-Sachs quantum cosmology}
	
	\author{Georgios Fanaras and Alexander Vilenkin}
	
	\address{ Institute of Cosmology, Department of Physics and Astronomy,\\ 
		Tufts University, Medford, MA 02155, USA}
	
	\begin{abstract}
		
		{We use a path-integral approach to study the tunneling wave function in quantum cosmology with spatial topology $S^{1}\times S^{2}$ and positive cosmological constant (the Kantowski-Sachs model).  
			If the initial scale factors of both $S^1$ and $S^2$ are set equal to zero, the wave function describes (semiclassically) a universe originating at a singularity.  This may be interpreted as indicating that an $S^1\times S^2$ universe cannot nucleate out of nothing in a non-singular way.  Here we explore an alternative suggestion by Halliwell and Louko that creation from nothing corresponds in this model to setting the initial volume to zero.  We find that the only acceptable version of this proposal is to fix the radius of $S^1$ to zero, supplementing this with the condition of smooth closure (absence of a conical singularity).  The resulting wave function predicts an inflating universe of high anisotropy, which however becomes locally isotropic at late times.}
		Unlike the de Sitter model, the total nucleation probability is not exponentially suppressed, unless a Gauss-Bonnet term is added to the action.
		
	\end{abstract}

	\maketitle

	\tableofcontents
	
	\section{Introduction}
	
	In quantum cosmology the entire universe is treated quantum mechanically and is described by a wave function, rather than by a classical spacetime. The wave function $\Psi(g,\phi)$ is defined on the space of all 3-geometries ($g$) and matter field configurations ($\phi$), called superspace.  It can be found by solving the Wheeler-DeWitt (WDW) equation \cite{DeWitt:1967yk}
	\beq
	{\cal H}\Psi=0,
	\eeq
	where ${\cal H}$ is the Hamiltonian operator.  Alternatively, one can consider the transition amplitude from the initial state $(g',\phi')$ to the final state $(g,\phi)$, which can be expressed as a path integral,
	\beq
	G(g,\phi|g',\phi') = \int_{(g',\phi')}^{(g,\phi)} {\cal D}g~{\cal D}\phi ~e^{iS},
	\label{pathint}
	\eeq
	where $S$ is the action.  In general, $G$ is a Green's function of the WDW equation \cite{Teitelboim}.  
	But if $(g',\phi')$ is at the boundary of superspace, then $G(g,\phi|g',\phi')$ is a solution of the WDW equation everywhere in the bulk of superspace, and the path integral (\ref{pathint}) may be used to define a wave function of the universe.
	
	The choice of the boundary conditions for the WDW equation and of the class of paths included in the path integral representation of $\Psi$ has been a subject of ongoing debate. The most developed proposals in this regard are the Hartle-Hawking (HH) \cite{Hartle:1983ai} and the tunneling \cite{Vilenkin:1986cy,Vilenkin:1987kf,Vilenkin:1994rn} wave functions.\footnote{For early work closely related to the tunneling proposal, see Refs.~\cite{Vilenkin:1982de,Linde:1983mx,Rubakov:1984bh,Vilenkin:1984wp,Zeldovich:1984vk}.}
	The intuition behind both of these proposals is that the universe originates `out of nothing' in a nonsingular way.  But despite a large amount of work, a consensus on the precise definition of these wave functions has not yet been reached.
	
	
	The two proposals have been thoroughly studied in the simple minisuperspace de Sitter model with $S^{3}$ spatial topology, where the only degree of freedom is the radius of the universe. A number of more complicated models with two or more degrees of freedom have also been considered.  Among them is the 
	Kantowski-Sachs (KS) model \cite{KS} which describes an anisotropic universe of spatial topology $S^{1}\times S^{2}$ with different scale factors.  
	We have recently presented a detailed analysis of the HH wave function in the KS model \cite{FV}, and our goal in the present paper is to extend this analysis to the tunneling wave function.

	We shall conclude this Introduction with some comments about prior work on this topic.\footnote{For earlier work on KS quantum cosmology see \cite{Laflamme,Louko:1988bk,Conradi,Louko:1988ia}.}
	Conti and Hertog \cite{CH} considered the semiclassical wave function by studying complex Schwarzschild-de Sitter instantons of the model, imposing boundary conditions suitable for a smooth closure of the 4-geometry.
	An advantage of this approach is that it is straightforward to obtain approximate expressions for the saddles by algebraic means. On the other hand, it does not allow one to rigorously define a convergent path integral and select which of these saddles contribute to the wave function. 
	Specifically, the divergence of the tunneling wave function found by CH can be attributed to the inclusion of saddle geometries that should not contribute. 
	
	Halliwell and Louko \cite{HL} used methods similar to Picard-Lefschetz theory 
	in order to define a steepest descent contour that renders the path integral convergent. With this approach it is usually the case that not all extremum geometries contribute to the integral, since the  lapse integration contour in the complex plane does not pass through all the saddles. In this paper we will follow this procedure, but we will have to extend the analysis to a non-vanishing cosmological constant, which makes the problem significantly more complicated.
	
	This paper is organized as follows. In Section 2 we review the classical dynamics of the KS model and its canonical quantization.
	The definition of the tunneling wave function is discussed in Section 3, where we outline the approach based on the outgoing wave condition in superspace, as well as the path integral definition.  In this paper we adopt the path integral approach and study
	alternative choices of boundary conditions for the path integral in Sections 4 and 5.  In Sec.4 we fix the scale factors of $S^1$ and $S^2$ on both initial and final spacetime boundaries.  The path integral in this case can be calculated exactly \cite{HL}.
	We find however that the resulting wave function is singular and thus is not an acceptable solution of the WDW equation.
	
	Section 5 is the main part of this work.  Here we investigate the choice of boundary conditions suggested by Halliwell and Louko \cite{HL}: we fix the two scale factors on the final boundary and require a smooth closure of the 4-geometry at the initial boundary.  In this case the path integral cannot be computed exactly, so we employ the methods of Picard-Lefschetz theory in order to make the integral over the lapse $N$ absolutely convergent. This is done by finding saddle points of the action in the complex plane and deforming the initial integration contour to a steepest descent path through the contributing saddles. 
	The wave function is then found in the WKB approximation by carrying out the Gaussian integrals in the vicinity of the saddles. 
	Since our analysis is approximate, we singled out three regimes of interest. Denoting the scale factors of $S^1$ and $S^2$ by $a$ and $b$ respectively and the cosmological constant by $\Lambda =H^2/3 \ll 1$ (in Planck units), we
	first looked into the case $b\approx1/H$.  This is the value at which the HH wave function gives a maximum probability \cite{FV}.  We found that the tunneling wave function exhibits an opposite behavior: the probability grows away from this value.
	We continued by probing the region of large $S^{2}$ 
	and found that the wave function is peaked at high anisotropy: $a/b\lesssim H\ll 1$. We note that in this regime CH found a divergence in the tunneling solution which we do not observe. Finally, we obtained the wave function in the limit of $a\ll 1/H$.  This wave function does not exhibit any exponential suppression as a function of $b$.
	Thus, the familiar picture of tunneling through a potential barrier does not hold in the KS model.
	
	Our results are summarized and discussed in Section 6.

	\section{Kantowski-Sachs model}
	
	\subsection{Classical dynamics}

	The Kantowski-Sachs (KS) model describing a homogeneous universe with spatial sections of $S^{1}\times S^{2}$ topology is represented by a Lorentzian metric as
	\beq
	ds^2 =- N^2 dt^2+a^2(t)dx^2+b^2(t) d\Omega^2 .
	\label{KSm}
	\eeq
	The Einstein-Hilbert action is then  
	\beq
	S= -\pi\int dt \left[\frac{1}{N}\left(a{\dot b}^2+2b{\dot a}{\dot b}\right)+Na(H^2 b^2-1)\right]+S_{b}.
	\label{S1}
	\eeq
	where the integration is carried out from the  initial boundary $\mathcal{B}_{0}$ to the  final boundary $\mathcal{B}$. The inclusion of the boundary term is needed if one chooses to impose Neumann conditions on one of the scale factors at $\mathcal{B}_{0}$ and is given by
	\beq
	S_{b}=-\left[\frac{\pi}{N}\frac{d}{dt}(ab^{2})\right]_{\mathcal{B}_{0}}.
	\label{Sb}
	\eeq
	
	Varying the action with respect to the scale factors $a$, $b$ we obtain the classical equations of motion in the gauge $N=1$:
	\beq
	\ddot{a}b+a\ddot{b}+\dot{a}\dot{b}-H^{2}ab=0,
	\label{eom1}
	\eeq
	\beq
	2b\ddot{b}+\dot{b}^{2}+1-H^{2}b^{2}=0,
	\label{eom2}
	\eeq
	and varying with respect to the lapse $N$ we obtain the Hamiltonian constraint:
	\beq
	\dot{b}^{2}+2\frac{\dot{a}}{a}\dot{b}b+1-H^{2}b^{2}=0.
	\label{hc}
	\eeq
	Combining equations (\ref{eom2}) and (\ref{hc}) we obtain
	\beq
	\dot{b}=\kappa a,
	\eeq
	where $\kappa$ is an integration constant. Plugging back into equation (\ref{eom2}) and integrating we obtain
	\beq
	\kappa a=\dot{b}=\pm\sqrt{\frac{H^{2}b^{2}}{3}-1+\frac{2M}{b}},
	\label{sol2}
	\eeq
	where again $M$ is an integration constant. 
	
	The above solution corresponds  to a Euclidean Schwarzschild-deSitter black hole of mass M. 
	In Euclidean time $t_{E}=it$ and in the gauge $N=1$ the metric (\ref{KSm}) becomes
	\beq
	ds_{E}^{2}=\left(1-\frac{2M}{b}-\frac{H^{2}b^{2}}{3}\right)d\lambda^{2}+\frac{db^{2}}{\left(1-\frac{2M}{b}-\frac{H^{2}b^{2}}{3}\right)}+b^{2}d\Omega_{2}^{2},
	\eeq
	where $\lambda=\kappa x$. 
	The black hole mass can be expressed in terms of the boundary data. Setting $a(t_{E}=0)=a'$ and $b(t_{E}=0)=b'$ we have
	\beq
	M=\frac{b'}{2}\left(1-\frac{H^{2}b'^{2}}{3}+\kappa^{2}a'^{2}\right).
	\label{mass}
	\eeq
	
	The limiting case in which $\kappa=0$ corresponds to the Nariai solution \cite{Nariai} {in which}:
	\beq
	b=\frac{1}{H} ~~,~~~ \ddot{a}=H^{2}a ~~,~~~ M=\frac{1}{3H} ~.
	\label{sol1}
	\eeq
	It describes a 4-geometry $dS_{2}\times S^{2}$, where the characteristic radius of both $dS_2$ and $S^2$ is $H^{-1}$.

	\subsection{The WdW equation}
	
	The WdW equation of the KS model can be more conveniently realized by switching to time variable $d\tau=a(t)dt$. In this representation the metric is
	\beq
	ds^2=-\frac{N^2}{a^2}d\tau^2+a^2 dx^2+b^2 d\Omega^2,
	\label{KSEmetric}
	\eeq
	where $N$, $a$ and $b$ are functions of time $\tau$, which we can choose to vary in the range $0<\tau<1$.  After substituting this in the Lorentzian Einstein-Hilbert action  and integrating over $x$ and over the angular variables, the action reduces to
	\beq
	S=-\pi\int_0^1 d\tau \left[\frac{{\dot b}{\dot c}}{N}+N(H^2 b^2-1)\right],
	\eeq
	where we have introduced a new variable $c=a^2 b$.

	The constraint equation is obtained by varying the action with respect to $N$:
	\beq
	\frac{{\dot b}{\dot c}}{N}-N(H^2 b^2-1)=0,
	\label{3}
	\eeq
	where overdots stand for derivatives with respect to $\tau$.
	
	The momenta conjugate to the variables $b$ and $c$ are
	\beq
	p_b=-\pi{\dot c}/N,~~~~ p_c=-\pi{\dot b}/N.
	\eeq
	Using this in the constraint equation (\ref{3}) and replacing $p_b\to-i\p/\p b$, $p_c\to -i\p/\p c$, we obtain the WDW equation 
	\beq
	\pi{\cal H}\Psi=\left[\p_b\p_c+\pi^2(H^2 b^2-1)\right]\Psi=0.
	\label{KSWDW}
	\eeq
	
	This can be rewritten in the form of a Klein-Gordon (KG) equation,
	\beq
	\frac{1}{\sqrt{-f}}\p_\alpha\left(\sqrt{-f}f^{\alpha\beta}\p_\beta \Psi\right) +V\Psi=0,  
	\eeq
	where the potential is $V=\pi^2(H^2 b^2 -1)$, $f_{\alpha\beta}$ is the minisuperspace metric, and $f=\det(f_{\alpha\beta})$.  With $b$ and $c=a^2 b$ used as coordinates, the minisuperspace metric is given by $f_{bb}=f_{cc}=0$, $f_{bc}=f_{cb}=2$, and $\sqrt{-f}=2$.  The contravariant components of the metric are $f^{bb}=f^{cc}=0$ and $f^{bc}=f^{cb}=1/2$.  The conserved current density corresponding to this KG equation is
	\beq
	j^{\alpha}=i\sqrt{-f}f^{\alpha\beta}(\Psi^{*}\p_{\beta}\Psi-\Psi\p_{\beta}\Psi^{*}).
	\label{jalpha}
	\eeq
	It satisfies
	\beq
	\p_\alpha j^\alpha=0,
	\eeq
	or
	\beq
	\frac{\p j^b}{\p b}+\frac{\p j^c}{\p c}=0.
	\eeq
	This suggests that
	\beq
	dP_b = j^c db
	\label{Pb}
	\eeq
	can be interpreted as the probability distribution for $b$ at a fixed value of $c$; then $c$ plays the role of a "clock" variable.  Similarly,
	\beq
	dP_c = j^b dc
	\label{Pc}
	\eeq
	can be interpreted as the probability distribution for $c$ at a fixed value of $b$, with $b$ being the clock variable \cite{Vilenkin:1988yd}.

	\section{Tunneling wave function}
	
	The tunneling wave function $\Psi_{T}$ was originally defined by imposing a boundary condition
	in superspace \cite{Vilenkin:1986cy}. The tunneling boundary condition states that $\Psi_{T}$ includes only outgoing
	modes, with the probability flux directed toward the boundary, at singular boundaries of
	superspace.\footnote{The definition of outgoing modes in quantum cosmology is discussed in Ref.\cite{Vilenkin:1994rn}.}
	The division of the superspace boundary into regular and singular parts has not been specified in the general case; here is a somewhat heuristic prescription.
	The boundary of superspace can be thought of as consisting of singular configurations
	which have some regions of {infinite} 3-curvature or {infinite} matter density, as well as configurations of infinite 3-volume and infinite values of matter fields. The regular part of the
	boundary includes singular 3-geometries which can be obtained by slicing regular Euclidean
	4-geometries. For example, if a 4-sphere embedded in a $5D$ Euclidean space is sliced by parallel planes, one gets 3-spheres
	of vanishing radius and infinite curvature at the two poles, even though the 4-geometry is
	perfectly regular there.
	The outgoing wave boundary condition is sometimes supplemented by requiring that the wave function is normalizable or that its modulus is bounded from above.
	
	In the simplest minisuperspace model, describing a de Sitter universe, the wave function depends only on the scale factor $a$ and the superspace is a half-line,
	$0 \leq a < \infty$. The regular boundary in this case is at $a = 0$ and the singular boundary
	is at $a=\infty$. The general picture is that the probability flux is injected into superspace
	through the regular boundary and flows out through the singular boundary.
	
\begin{figure}[h!]
		\includegraphics[scale=0.4]{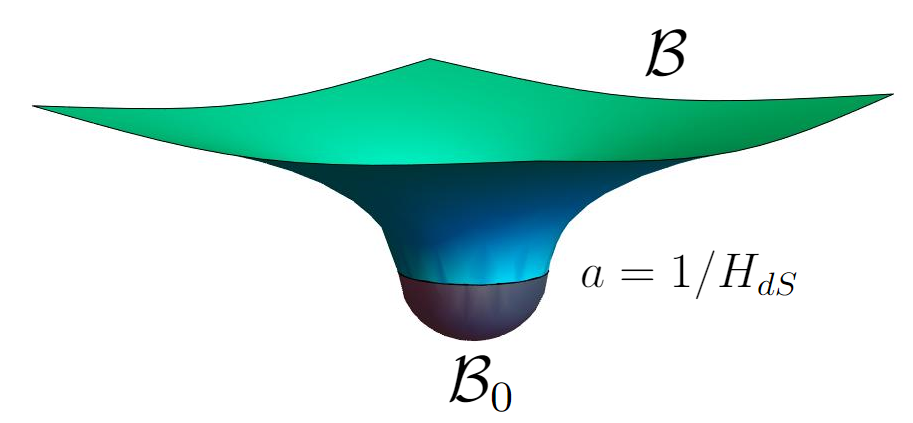}
		\caption{Nucleation of a de Sitter universe. 
			Half of de Sitter space is matched to Euclidean hemisphere at the bounce radius $a=1/H_{dS}$.}
		\label{fig1}
		
	\end{figure}

	The origin of the universe in the de Sitter model can be pictured semiclassically as illustrated in Fig.\ref{fig1}.  The purple hyperboloid at the top is the classical de Sitter space and the blue hemisphere at the bottom is its Euclidean continuation.  Such a continuation is necessary because Lorentzian geometries cannot close off at the bottom without a singularity.  For this reason the regular boundary of superspace is specified in terms of Euclidean geometries. 
	
	The classical de Sitter model describes a universe contracting from infinite size, bouncing at the turnaround radius $a=1/H_{dS}$, and re-expanding.  (Here $H_{dS}$ is the de Sitter expansion rate.)
	The tunneling wave function represents a universe that transits from $a=0$ through the classically forbidden range $0<a<1/H_{dS}$ and expands from there.  It is formally similar to a wave function describing quantum tunneling through a potential barrier.
	
	It should be noted that the analogy between quantum cosmology and quantum tunneling often breaks down in models with more than one degree of freedom.  Consider for example a model with coordinates $q^\alpha$ and conjugate momenta $p_\alpha$, satisfying the Hamiltonian constraint ${\cal H}= K(q,p)+V(q)=0$, where the kinetic energy $K$ is a non-negative-definite quadratic form in momenta.  Then classically we have $K\geq 0$, so the range $V(q)>0$ is classically forbidden.  On the other hand, in minisuperspace cosmology the quadratic form $K$ is not generally positive-definite and not even bounded from below -- not even at the classical level.  Hence, in multi-dimensional models superspace cannot generally be divided into classically allowed and classically forbidden regions.

	An alternative definition of the tunneling wave function has been given in terms of a
	path integral \cite{Vilenkin:1994rn}. It states that $\Psi_{T}(g)$ is given by a path integral over Lorentzian histories
	interpolating between a 3-geometry of vanishing size (”nothing”) and a given configuration $g$, with the lapse integration taken
	over the positive range $0<N<\infty$.  As discussed in \cite{Teitelboim}, a lapse integral
	over a half-infinite range gives a Green’s function of the WDW equation. But in this case
	the source term has support only at geometries of vanishing size, which are at the boundary of superspace. Hence one can expect that $\Psi_{T}$ is a solution of the WDW equation everywhere in the bulk of superspace. 
	The positive lapse condition can be thought of as a causality requirement \cite{Teitelboim}: the histories included in the path integral are to the future of the origin event of ”nothing”.  In the present paper we shall adopt the path integral approach to the tunneling wave function.

	In the case of de Sitter model the path integral is taken over the scale factor $a(t)$ with the boundary condition $a(0)=0$.  We note that this boundary condition, combined with the classical constraint equation ${\dot a}^2+1=H_{dS}^2a^2$ implies ${\dot a}(0)=\pm i$.  
	After Euclidean continuation $t=i\tau$, this gives $da/d\tau=\pm 1$, which is the condition of a smooth closure (absence of a conical singularity) in the Euclidean geometry at $\tau=0$.  Thus the boundary condition $a=0$ enforces the regularity condition at a semiclassical level. 
	Alternatively, one could impose the regularity requirement ${da/d\tau}(0)=\pm 1$ as a boundary condition.  Then the constraint equation would imply semiclassical closure, $a(0)=0$.
	The Lorentzian path integral for the de Sitter model with both of these boundary conditions has been calculated in Refs.\cite{Halliwell:1988ik,Feldbrugge:2017kzv}. In both cases the result coincides with the wave function obtained from the outgoing-wave boundary condition.  

	
	Turning now to the KS model, we first need to 
	specify the class of histories included in the path integral.  By analogy with the de Sitter model, one might consider histories originating from a configuration of vanishing 3-geometry, $a=b=0$.  However, it has been shown in Ref.\cite{HL} that Euclidean 4-geometries admitting $S^1\times S^2$ slicing with radii $a$ and $b$ necessarily have a divergent 4-curvature in the limit $a,b\to 0$ and are therefore singular even at the semiclassical level.  
	
	The conclusion could be that a universe of topology $S^1\times S^2$ cannot be created from nothing.  In this paper we shall explore an alternative idea, suggested by Halliwell and Louko in Ref.\cite{HL}.
	We shall relax the condition $a=b=0$ and require that only one of the two scale factors,
	$a$ or $b$, is equal to zero.  One possibility is then to fix the other scale factor at a nonzero value that is consistent with a non-singular geometry.  Alternatively, we can leave the other scale factor unspecified and impose a regularity condition excluding conical singularities, as we mentioned for the de Sitter model.  We shall discuss both of these approaches here.



	\section{Fixing initial scale factors}
	
	The transition amplitude from the initial state $\{a',b'\}$ to the final state $\{a,b\}$ in the KS model has been calculated by Halliwell and Louko in Ref.\cite{HL}.  It is given by 
	\beq
	G(a,b|a',b')=\int_0^\infty \frac{dN}{N}\exp\left[i\pi\left(\alpha N-\frac{\beta}{N}\right)\right],
	\label{GN}
	\eeq
	where 
	\beq
	\alpha=1-\frac{H^2}{3}(b^2+bb'+{b'}^2),~~~ \beta= (a^{2}b-a'^{2}b')(b-b').
	\label{alphabeta}
	\eeq
	The contour of $N$-integration is generally complex; here we choose it to lie along the positive real axis, as required for the tunneling wave function.  
	
	The integral over $N$ in (\ref{GN}) can be expressed in terms of Bessel functions \cite{HL}.
	The resulting wave function is
	\beq
	\Psi_{T}=-i\pi H_{0}^{(2)}(2\pi(-X)^{1/2})
	\eeq
	for $X<0$ and
	\beq
	\Psi_{T}=2K_{0}(2\pi(X)^{1/2})
	\eeq
	for $X>0$, where $X=\alpha\beta$.
	If we set $a'=0$, then
	\beq
	-X= a^2 b^2\left({\frac{H^{2}b^{2}}{3}-1+\frac{2M}{b}}\right)
	\eeq
	with $M$ from Eq.(\ref{mass}), and if we set $b'=0$, then $M=0$ and $X$ is independent of $a'$, 
	\beq
	-X= a^2 b^2\left({\frac{H^{2}b^{2}}{3}-1}\right).
	\eeq 
	We note that the same value of $M=0$ is obtained for $a'=0,~b'=H^{-1}\sqrt{3}$; hence this choice of parameters gives the same wave function as $b'=0$ with arbitrary $a'$.
	
	Let us now consider the wave function with $M=0$.  For $Hb>\sqrt{3}$ it is given by
	\beq
	\Psi_{T}(Hb>\sqrt{3})=-i\pi H_{0}^{(2)}\left(2\pi ab\sqrt{\frac{H^{2}b^{2}}{3}-1}\right).
	\label{fixscale}
	\eeq
	To verify that this wave function satisfies the outgoing flux criterion, we first note its asymptotic form for large argument:
	\beq
	\Psi_{T}\propto\exp\left[-2i\pi ab\sqrt{\frac{H^{2}b^{2}}{3}-1}\right],
	\label{asymptfixed}
	\eeq
	where we have ignored the pre-exponential factor.  This gives a good approximation for the exponent, as long as $b$ is not very close to $H^{-1}\sqrt{3}$.  Acting with the momentum operators and using the gauge $N=1$ we obtain:
	\beq
	\Pi_{a}\Psi_{T}=-i\frac{\p\Psi_{T}}{\p a}\longrightarrow\dot{b}=+\sqrt{\frac{H^{2}b^{2}}{3}-1}>0
	\eeq
	\beq
	\Pi_{b}\Psi_{T}=-i\frac{\p\Psi_{T}}{\p b}\longrightarrow\frac{\dot{a}}{a}=\frac{bH^{2}}{3\sqrt{\frac{H^{2}b^{2}}{3}-1}}>0.
	\eeq
	The solution of these equations for $b$ and $a$ is
	\beq
	b(t)=\frac{\sqrt{3}}{H}\cosh\left(\frac{Ht}{\sqrt{3}}\right),~~~a(t)=D\sinh\left(\frac{Ht}{\sqrt{3}}\right),
	\label{dS}
	\eeq
	where $D>0$ is a constant parameter.  
	
	In the semiclassical approximation, the wave function (\ref{fixscale}) describes a congruence of expanding classical trajectories (\ref{dS}) with different values of $D$.  The trajectories start at $a'=0$, $b'=H^{-1}\sqrt{3}$ and extend to $a,b\to\infty$.
	The trajectory with $D=H^{-1}\sqrt{3}$ describes a de Sitter space with expansion rate $H/\sqrt{3}$, while for other values of $D$ the geometries (\ref{dS}) have conical singularities at $t=0$. For large values of $Ha$ and $Hb$ all these geometries approach an expanding de Sitter space.
	
	
	
	The wave function for $X>0$ can be similarly analyzed.  The resulting congruence of trajectories is a Euclidean continuation of (\ref{dS}):
	\beq
	b(\tau)=\frac{\sqrt{3}}{H}\cos\left(\frac{H\tau}{\sqrt{3}}\right),~~~a(\tau)=iD\sin\left(\frac{H\tau}{\sqrt{3}}\right),
	\label{dSE}
	\eeq
	where $\tau$ is the Euclidean time.  It describes trajectories staring at $a'=0$, $b'=H^{-1}\sqrt{3}$ and ending at $a=D$, $b=0$.  Once again, the trajectories with $D\neq H^{-1}\sqrt{3}$ have conical singularities at $\tau=0$.
	We note that even though the wave function (\ref{fixscale}) is obtained for several choices of the initial values $a',b'$, the congruence of trajectories that it describes corresponds to only one of these choices: $a'=0, ~b'=H^{-1}\sqrt{3}$.
	
	We can use the conserved current (\ref{jalpha}) to find the probability distribution for $c=a^2 b$ at a fixed value of $b$, with $b$ playing the role of a clock.  Using the Wronskian products of the Hankel functions, we find that the current is given by
	\beq
	j^{\alpha}=4\pi \sqrt{-f}f^{\alpha\beta}\frac{\p \ln(X)}{\p y^{\beta}},
	\label{curr}
	\eeq
	where $y^\beta$ are the superspace coordinates $b,c$ and 
	$X=2\pi \sqrt{cb\left(\frac{H^{2}b^{2}}{3}-1\right)}$. From Eq.(\ref{Pc}) we obtain the probability distribution \beq
	dP_c\propto j^b dc \propto \frac{dc}{c}.
	\label{dPc}
	\eeq
	Since $b$ is fixed, the distribution for $a$ is $dP_a\propto da/a$.  This distribution is not normalizable, but it admits a simple interpretation: values of $a$ in each logarithmic interval are equally probable (at any given value of $b$).
	
	The problem with the wave function (\ref{fixscale}) is that it exhibits a logarithmic singularity at $Hb=\sqrt{3}$.  Hence it is not a solution of the WDW equation in the entire superspace, $0\leq a,b<\infty$.
	Furthermore, the semiclassical geodesic congruence described by this wave function has conical singularities and thus does not correspond to a non-singular origin of the universe.  The reason is that the boundary conditions that we used for initial scale factors do not enforce regular geometry even at the semiclassical level.
	Similar features are obtained for wave functions specified by a vanishing initial scale factor $a'=0$ with an arbitrary value of $b'$.  We therefore conclude that this class of wave functions is not a suitable choice for the tunneling wave function of the universe.

	\section{Smooth closure}
	
	We now consider the boundary condition of a smooth closure of Euclidean geometry with one of the scale factors vanishing.  In the rest of the paper it will be more convenient to switch to Euclidean signature, by replacing $N\to -iN$. 
	
	It has been shown in \cite{HL} that in a classical Euclidean KS geometry a smooth closure can be achieved by one of the following two sets of boundary conditions:
	\beq
	a'=0,~~\frac{1}{N}{\dot a}'=\pm 1, ~~ \frac{1}{N}{\dot b}'=0,
	\label{classbc1}
	\eeq
	\beq
	b'=0,~~\frac{1}{N}{\dot b}'=\pm 1, ~~ \frac{1}{N}{\dot a}'=0,
	\label{classbc2}
	\eeq
	where a prime indicates evaluation at the  initial boundary and the Euclidean lapse parameter $N$ is now real.
	The former set corresponds to a smooth closing of $S^{1}$, while the latter to a smooth closing of $S^{2}$. 
	
	Neither set of boundary conditions can be implemented in quantum theory.  This becomes apparent if we
	introduce new variables
	\beq
	A=b^{2},~~B=ab
	\eeq
	with conjugate momenta
	\beq
	P_A=-\frac{\dot a}{2N},~~P_B=-\frac{\dot b}{N}.
	\eeq
	{In this representation, the regularity conditions (\ref{classbc1}), (\ref{classbc2}) take the form}
	\beq
	B'=0,~~{P_A}'=\mp \frac{1}{2},~~P_{B}'=0
	\label{bc1}
	\eeq
	and
	\beq
	A'=0,~~{P_A}'=0,~~P_{B}'=\mp1.
	\label{bc2}
	\eeq
	
	{Quantum mechanically, however, one is not allowed to impose these regularity conditions in their entirety, since that would violate the uncertainty principle:  we cannot fix both a superspace variable and its conjugate momentum at the boundary.  The best we can do is to enforce two of the three conditions.  The third can then be inferred from the classical equations of motion, indicating that the semiclassical wave function would approximately describe a regular geometry.  As we shall see in the next subsection,  setting $b'=0$ does not allow one to specify the momenta, since they appear in the action factored to $b'$.  In fact, with $b'=0$ one necessarily gets the same wave function (\ref{fixscale}) as we discussed in Sec 4, which we have concluded should be disqualified. We therefore focus on the boundary conditions
		\beq
		B'=0,~~{P_A}'=\mp \frac{1}{2}.
		\label{bc1'}
		\eeq
		

		\subsection{General formalism}

		{The formalism for calculating the propagator with specified values of $a$ and $b$ at the future boundary $\mathcal{B}$ and boundary conditions (\ref{bc1'}) at the initial boundary $\mathcal{B}_{0}$ has been developed in \cite{HL,FV}. Here we will outline the approach for constructing the propagator;  the details can be found in these references.
			
			The fist step is to compute the action 
			by integrating Eq.(\ref{S1}) 
			while also evaluating the boundary term (\ref{Sb}). After substituting $a'=0$, the result is
			\beq
			S_{E}=\pi\left[ \frac{H^2}{3}N(b^2+bb'+{b'}^2)-N-\frac{a^2 b}{N}(b-b')+2b'^2{P_{A}}'\right].
			\label{SE0}
			\eeq
			This is the action for $a(t)$ and $b(t)$ satisfying the classical equations of motion and boundary conditions}, but not the constraint. 
		The quantity $b'$ in Eq.(\ref{SE0}) has to be expressed in terms of the boundary data $\{a,b\}$, $\{P_{A}'\}$ and $N$ using the equations of motion.  This gives
		\beq
		b'=-\frac{b}{2N}\frac{a^2+H^2 N^2/3}{{2P_{A}}'+H^2 N/3}.
		\label{b'b}
		\eeq
		Substituting this expression in the action (\ref{SE0}) and simplifying, we have
		\beq
		S_E=\frac{\pi N}{3}(H^2 b^2-3)-\frac{\pi}{N}a^2 b^2-\frac{\pi b^2}{4N^2}\frac{\left(a^2+\frac{H^2 N^2}{3}\right)^2}{2P_{A}'+\frac{H^2 N}{3}}
		\label{SE1}
		\eeq
		The transition amplitude from the initial state $Z^{\beta}=\{B',P_{A}'\}$ to the final state $Q^{\alpha}=\{A,B\}$ can now be expressed as
		\beq
		\Psi_{T}(Q^{\alpha})=\int_{C}\mu(Q^{\alpha},Z^{\beta},N) \exp\left[-S_{E}\left(Q^{\alpha};N|Z^{\beta}\right)\right]dN
		\label{amplitude}
		\eeq
		where $\mu$ is the semiclassical prefactor and $C$ is a Lorentzian integration contour along the positive imaginary axis.}
	
	At this point the transition amplitude is not yet fully defined since we have not specified which of the values of $P_{A}'=\pm1/2$ should be used. We make this choice by requiring that the integral (\ref{amplitude}) is convergent.  Representing $N=iy$ with $0<y<\infty$, let us examine the behavior of the integrand at $y\to 0$.
	In this limit the action (\ref{SE1}) becomes
	\beq
	S_{E}\approx\left(\frac{\pi b^{2}a^{4}}{8P_{A}'}\right)\frac{1}{y^{2}}
	\eeq
	In order for the integral of $\exp(-S_{E})$ to converge near the origin, the appropriate choice is $P_{A}'=+1/2$.
	This is opposite to the choice $P_{A}'=-1/2$ made in Refs.\cite{FV} for the Hartle-Hawking wave function.
	
	In the limit $y\rightarrow+\infty$ the action becomes
	\beq
	S_{E}\approx i\pi(H^{2}b^{2}-4)\frac{y}{4}.
	\eeq
	Thus the integral of $\exp(-S_{E})$ will be convergent in the following cases:
	\beq
	\arg(y)\in(-\frac{\pi}{2},0) \ \ \ , \ \ \ Hb>2
	\eeq
	or
	\beq
	\arg(y)\in(0,\frac{\pi}{2}) \  \ \ , \ \ \ Hb<2 .
	\eeq
	So, depending on the sign of $(Hb-2)$, the integration contour must be given an appropriate tilt in order to converge.
	
	Unlike in the case of transition between fixed scale factors, the amplitude (\ref{amplitude}) cannot be evaluated exactly.  We shall therefore compute the integral following the methods of Picard-Lefschetz, namely distorting the integration contour $C$ to a steepest descent/ascent path going through (at least) one of the saddle points of the action. 
	
	In order to simplify the analysis we will rescale our variables:
	\beq
	u=H^2 a^{2} \ \ , \ \ v=H^{2}b^{2} \ \ , \ \ \tilde{N}=H^2 N \ \ , \ \ \tilde{S_{E}}= \frac{H^2 S_{E}}{\pi}.
	\label{rescaled}
	\eeq
	In this representation the rescaled action becomes
	\beq
	\tilde{S}_{E}=\frac{ \tilde{N}(v-3)}{3}-\frac{uv}{\tilde{N}}-\frac{v}{4\tilde{N}^2}\frac{\left(u+\frac{\tilde{N}^2}{3}\right)^2}{2P_{A}'+\frac{\tilde{N}}{3}}.
	\label{RescaledSE}
	\eeq
	The saddle points of the action are found by solving the algebraic equation 
	\beq
	\frac{\p \tilde{S_{E}}}{\p \tilde{N}}=0.
	\label{saddleeq}
	\eeq
	This is a quintic equation for $\tilde{N}$ that cannot be solved exactly for all values of $u$, $v$, except in some limits. 

	{Once the steepest descent contour has been specified, along with the contributing saddle  points, the integral can be approximated by expanding the action about the extrema and carrying out a Gaussian approximation. Specifically, for each  contributing saddle $N_{i}$ the action can be expanded in the vicinity of $N_{i}$ as}
	\beq
	S_{E}(N-N_{i})\approx S_{E}(N_{i})+\frac{S_{NN}(N_{i})}{2}(N-N_{i})^{2} ,
	\eeq
	{where $S_{NN}=\p^{2}S_{E}/\p N^{2}$. Inserting in the integral (\ref{amplitude}) and integrating over $d(N-N_{i})$ we can approximate the transition amplitude as}
	\beq
	\Psi\propto\sum_{i} \frac{\mu(N_{i})}{\sqrt{S_{NN}(N_{i})}}\exp[-S_{E}(N_{i})].
	\label{WKBsol}
	\eeq
	{Thus, the overall WKB pre-exponential factor for each contributing saddle will be given by}
	\beq
	\tilde{\mu}=\frac{\mu}{\sqrt{S_{NN}}}.
	\label{WKBprefactor}
	\eeq
	
	{The semiclassical prefactor $\mu$ for the propagator (\ref{amplitude}) has been derived in []. For the choice of $P_{A}'=+1/2$ it is given by}
	\beq
	\mu\left(a,b,N\right)\propto \frac{b^{\prime}}{N\sqrt{H^{2}N+3}},
	\label{mu}
	\eeq
	where $b'$ is the initial radius of $S^{2}$ defined in (\ref{b'b}). The denominator of $\mu$ introduces a branch cut. For convenience, we can choose its orientation to be along the real axis at $H^{2}N<-3$, but as we will see this will not affect the calculation of the propagator, since any integration along the branch cut is exponentially suppressed.

	\subsection{$S^{2}$ of radius $Hb\approx1$}
	
	The classical KS model has a Nariai solution (\ref{sol1}), which describes a 4-geometry $dS_{2}\times S^{2}$ with the radius of $S^{2}$ being $b=1/H$.  The $dS_{2}$ part of the geometry is a circle $S^1$ undergoing inflation with an expansion rate $H$.  Our analysis of the Hartle-Hawking wave function $\Psi_{HH}$ for this model \cite{FV} showed that it gives a probability distribution for $b$ at fixed values of $a$ which is peaked at $b=H^{-1}$.  So this wave function predicts Nariai-type evolution as the most probable scenario.  To compare this prediction with that of the tunneling wave function $\Psi_T$, we shall now study the behavior of $\Psi_T$ in the regime $Hb\approx 1$.
	
	
	\subsubsection{Saddles and contours}
	
	Following the method of Ref.\cite{FV}, we shall first find the saddle points for $Hb=1$ and then consider small perturbations of those points.  Using the rescaled variables (\ref{rescaled}) and setting $v=H^2 b^2=1$, the action (\ref{RescaledSE}) becomes
	\beq
	{\tilde S}_{E0}=-\frac{2\tilde{N}}{3}-\frac{u}{ \tilde{N}}-\frac{\left(3u+\tilde{N}^{2}\right)^{2}}{12\tilde{N}^{2}\left(\tilde{N}+3\right)},
	\eeq
	where the subscript $0$ indicates zeroth order with respect to $(1-v)$. The saddle equation (\ref{saddleeq}) for $v=1$ is
	\beq
	\left(\tilde{N}^{2}+2\tilde{N}+u\right)\left(\tilde{N}^{3}+4\tilde{N}^{2}-3\tilde{N}u-6u\right)=0.
	\label{SSol}
	\eeq
	It has two solutions,
	\beq
	\tilde{N}_{1,2}=-1\pm\sqrt{1-u},
	\label{N12}
	\eeq
	which are real for $u\le 1$ and form a complex conjugate pair for $u>1$, along with three solutions $\tilde{N}_{3,4,5}$ which are real for all $u\ge0$ and whose explicit form will not be needed.
	
	{The saddle points and steepest descent/ascent lines for $Ha>1$ are shown in Fig.\ref{fig2} Our nearly Lorentzian contour can be distorted into a contour running from $N = 0$ along the arc through the saddle $N_{1}$, all
		the way to $N_{5}$.
		Then it takes a turn and runs along the negative $N$ axis to $N \rightarrow-\infty $.
		This contour is dominated by the saddle point at $N_{1}$.
		\begin{figure}[h!]
			\includegraphics[scale=0.5]{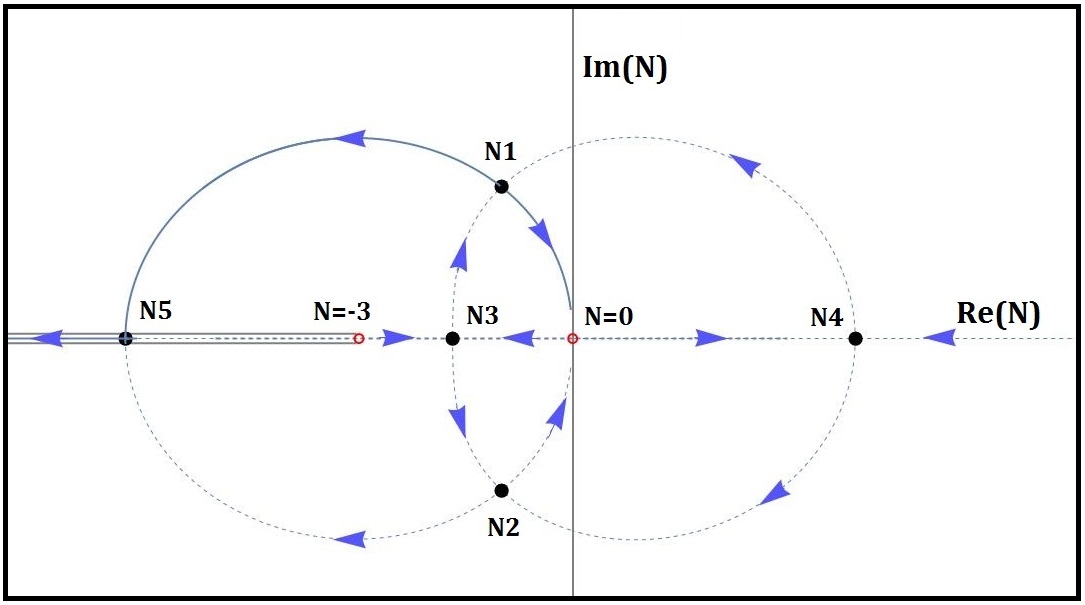}
			\caption{The steepest descent contours for $Ha>1$  and $Hb=1$.
				The arrowheads point to the direction where $Re(-\tilde{S}_{E})$ decreases. The saddles $\tilde{N}_{i}$ are marked with solid dots and the singularities with circles. Note the branch cut at $\tilde{N}\in(-\infty,-3)$. Our nearly-Lorentzian contour corresponds to the solid curve starting from the singularity $N=0$; it is dominated by the saddle $N_{1}$.}
			\label{fig2}
			
		\end{figure}

		We now consider small deviations from $Hb=1$.
		In the region $Ha>1$ we have identified the saddle $\tilde{N}_{1}$ as the dominant one, so we will introduce a shift $x$ defined by
		\beq
		\tilde{N}=\tilde{N}_{1}+x ,
		\label{Npert}
		\eeq
		where  $\tilde{N}_{1}$ is given by (\ref{N12}) and $|x|<<1$. We insert this into the action (\ref{RescaledSE}) and expand to second order in $x$:
		\beq
		\tilde{S}_{E}\approx1-i (1-v)\sqrt{u-1}-(1-v)x+\frac{v}{f(u)}x^{2}+O(x^{3}),
		\label{expansion}
		\eeq
		where
		\beq
		3f(u)=\left(2i+2\sqrt{u-1}-iu-iu^{2}\right)(u-1)^{-3/2}.
		\label{functionf}
		\eeq
		The action is extremized with
		\beq
		x=\frac{1-v}{2v}f(u).
		\label{x}
		\eeq
		
		To lowest order in $(1-Hb)$, we have $x\propto(1-Hb)$, so the contribution of the x-dependent terms to the action is ${\cal O}[(1-Hb)^{2}]$ and we can write  
		\beq
		S_E\approx \frac{\pi}{H^2}-\frac{2i\pi}{H^2}(1-Hb)\sqrt{H^2a^2-1}+{\cal O}[(1-Hb)^2].
		\label{SEab}
		\eeq
		The higher order correction term ${\cal O}[(1-Hb)^2]$ is proportional to $f(u)$ and will play an important role in the probability distribution that the tunneling wave function predicts.

		\subsubsection{WKB tunneling wave function}
		
		We are now in a position to compute the WKB tunneling wave function in the region $Hb\approx1$ and $Ha>1$ following the prescription (\ref{WKBsol}).
		The WKB prefactor $\tilde{\mu}$ defined in (\ref{WKBprefactor}) can be computed for the saddle $N_{1}$ as:
		\beq
		\tilde{\mu}\propto\frac{1}{\sqrt{H^{2}a^{2}-1}}
		\eeq
		Keeping only the lowest non-trivial orders of $(1-Hb)$ in the action (\ref{SEab}), we arrive at an expression for the tunneling wave function:
		\beq
		\Psi_{T}(Ha>1)\propto \frac{1}{\sqrt{H^{2}a^{2}-1}}\exp\left(-\frac{\pi}{H^{2}}\right)\exp\left(\frac{2\pi i}{H^2}(1-Hb)\sqrt{H^2a^2-1}\right).
		\label{PsiTHa>1v=1}
		\eeq
		
		There are a few things to note about this solution. It exhibits a WKB-type divergence at the turning point $Ha=1$, as expected. Additionally, the tunneling exponential suppression factor $\exp(-\pi/H^{2})$ is present. This is a consequence of the choice $P_{A}'=+1/2$ for our boundary condition. {Finally, it is easily verified that the wave function $\Psi_{T}(Ha>1, Hb=1)$ describes an outgoing wave at $Ha\gg 1$, and thus $\Psi_{T}$ satisfies the outgoing wave boundary condition in the region $Hb\approx1$.}


		In order to obtain a probability distribution for $b$ in the region $Hb\approx1$ we must include higher order corrections to the wave function (\ref{PsiTHa>1v=1}). Making use of the perturbed action (\ref{expansion}) and ignoring corrections to the prefactor, we arrive at
		\beq
		\Psi_{T}(Ha>1)\propto \frac{1}{\sqrt{H^{2}a^{2}-1}}\exp\left(-\frac{\pi}{H^{2}}\right)\exp\left[\frac{2\pi i}{H^2}(1-Hb)\sqrt{H^2a^2-1}+f(a)(Hb-1)^{2}\right],
		\label{PsiTHa>1v=1pert}
		\eeq
		where the function $f(a)$ is given by (\ref{functionf}). Utilizing the methods of section 2.2, we calculate the current density:
		\beq
		dP_b\propto j^a db \propto \frac{db}{\sqrt{H^2 a^2-1}}\exp\left[2(Hb-1)^2 {\rm Re} f(a)\right].
		\eeq
		This can be interpreted as the distribution for $b$ on surfaces of constant $a$. The real part of $f(a)$ is given by:
		\beq
		{\rm Re} f(a)=\frac{2\pi}{3H^2(H^2 a^2-1)}.
		\eeq
		Thus the probability distribution can be written explicitly as:
		\beq
		dP\propto  \frac{db}{\sqrt{H^2 a^2-1}}\exp\left[\frac{4\pi(Hb-1)^2}{3H^2(H^2 a^2-1)}\right].
		\label{P}
		\eeq
		This distribution grows as we move away from $Hb=1$, favoring large values of $Hb$. 
		This is in contrast with the HH state for which the probability distribution is peaked at $Hb=1$ \cite{FV}. 

		\subsection{Large $S^{2}$ region $Hb\gg1$}
		
		
		\subsubsection{Saddles and contours}
		
		Since the distribution (\ref{P}) appears to favor large values of $Hb$, we shall now explore the behavior of the wave function at $Hb\gg 1$.
		In order to find the saddle points in this regime, we neglect the $-3$ term in the first parenthesis of (\ref{RescaledSE}). 
		Then the saddle equation $(\ref{saddleeq})$ can be factored:
		\beq
		\left(\tilde{N}^{2}+3u\right)\left(\tilde{N}^{3}+6\tilde{N}^{2}+12\tilde{N}+3u\tilde{N}+6u\right)=0.
		\eeq
		The corresponding solutions include two complex conjugate pairs and one real saddle. Two obvious solutions are
		\beq
		\tilde{N}_{1,2}=\pm i\sqrt{3u}
		\label{tildeN12}
		\eeq
		and we label the other pair as $\tilde{N}_{4,5}$ and the real saddle as $\tilde{N}_{3}$. {We note that the numbering of saddle points here is not related to the one used in Section 5.2}. 
		

		The steepest descent contours for this set of saddles are shown in Fig.\ref{fig3}. A nearly Lorentzian contour can be distorted into a contour that follows the steepest ascent path from the origin $N=0$ to the saddle $N_{1}$ and continues on the steepest descent path all the way to infinity in the first quadrant. The contour can be closed with an infinite arc at $N\rightarrow+i\infty$. The dominant contribution to the wave function comes from the saddle $N_{1}$.
		
		\begin{figure}[h!]
			\centering
			\includegraphics[scale=0.4]{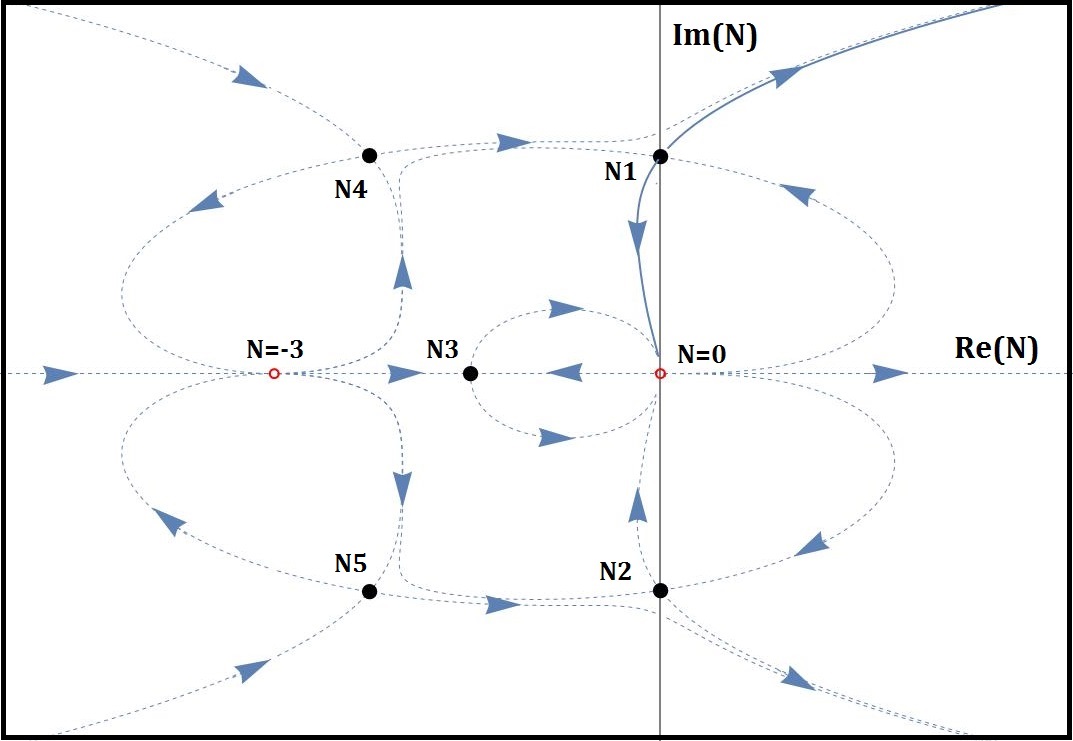}
			\caption{The steepest descent contours for $Hb\gg1$. The tunneling contour corresponds to the solid curve starting from the singularity $N=0$ and dominated by saddle $N_{1}$.}
			\label{fig3}
			
		\end{figure}

		To obtain a more accurate expression for the saddle point, we set 
		$\tilde{N}=\tilde{N}_{1}+x$ with $\tilde{N}_{1}$ from (\ref{tildeN12}) and $x\ll \tilde{N}_{1}$.  We then substitute it in the action (\ref{RescaledSE}), without neglecting the $-3$ term and expand the action to second order in $x$:
		\beq
		\tilde{S}_{E}\approx\frac{i\sqrt{u}(2v-3)}{\sqrt{3}}-x-\frac{vx^{2}}{3(u-i\sqrt{3u})}+\mathcal{O}(x^{3}).
		\eeq
		Extremizing with respect to $x$ we find
		\beq
		x=\frac{3(i\sqrt{3u}-u)}{2v} .
		\eeq
		This approximation is valid in the region of superspace where $v\gg1$ and $u\ll v^{2}$ in order for the condition $\tilde{N}_{1}\gg x$ to be satisfied. The $x$-dependent terms in the action introduce corrections $\mathcal{O}(v^{-1})$.  Switching back to variables $a$ and $b$, the perturbed action can be expressed as
		\beq
		S_{E}\approx\frac{2i\pi ab^{2}H}{\sqrt{3}}-\frac{\sqrt{3}i\pi a}{H}-\frac{3\sqrt{3}i\pi a}{4H^{3}b^{2}}+\frac{3\pi a^{2}}{4H^{2}b^{2}}+\mathcal{O}\left(\frac{1}{H^{3}b^{3}}\right).
		\label{correction Hb>>1}
		\eeq

		As with the case of $Hb\approx1$, the higher order correction terms will be important for obtaining the probability distribution that the tunneling wave function predicts.
		
		\subsubsection{WKB tunneling wave function }
		
		To lowest order in $(Hb)^{-1}$ the prefactor of the WKB wave function is:
		\beq
		\tilde{\mu}\propto \frac{\sqrt{a}}{b^{2}}e^{i\frac{\pi}{4}}.
		\eeq
		Thus to the leading order of WKB approximation 
		the tunneling wave function is given by
		\beq
		\Psi_{T}(Hb\gg1)\propto\frac{\sqrt{a}}{b^{2}}\exp\left[-\frac{2i\pi ab^{2}H}{\sqrt{3}}
		\right],
		\label{PsiT Hb>>1}
		\eeq
		where we have kept only the dominant contribution in $Hb\gg1$. It is straightforward to show that the above solution describes expanding asymptotically de Sitter universes satisfying 
		\beq
		{\dot b}/b\approx {\dot a}/a\approx H/ \sqrt{3} ,
		\label{approxdS}
		\eeq 
		just as in the case of fixed initial scale factors. Thus the outgoing wave criterion is satisfied.
		
		To find the probability distribution for the scale factors, we have to include higher order corrections to the action, as given by Eq.(\ref{correction Hb>>1}).
		Ignoring corrections to the prefactor, we have
		\beq
		\Psi_{T}(Hb\gg1)\propto\frac{\sqrt{a}}{b^{2}}\exp\left[-\frac{2i\pi ab^{2}H}{\sqrt{3}}+\frac{\sqrt{3}i\pi a}{H}+\frac{3\sqrt{3}i\pi a}{4H^{3}b^{2}}-\frac{3\pi a^{2}}{4H^{2}b^{2}}.
		\right]
		\eeq
		Noticing that the first three terms are the expansion of a square root, we can tidy this result to
		\beq
		\Psi_{T}(Hb\gg1)\propto\frac{\sqrt{a}}{b^{2}}\exp\left[-2i\pi ab\sqrt{\frac{H^{2}b^{2}}{3}-1}-\frac{3\pi a^{2}}{4H^{2}b^{2}}
		\right],
		\label{PsiTcompact}
		\eeq
		which is valid up to order $(Hb)^{-2}$. 
		This wave function has the same asymptotic form as the transition amplitude (\ref{asymptfixed}) that we obtained for fixed initial scale factors,
		the only difference being its amplitude, which is controlled by the last term in the bracket. We will show that this term is responsible for the predictions of the tunneling wave function.
		
		Following the formalism of Section 2.2 we calculate the probability current on surfaces of constant $a$. The calculation is simplified if we use the compact form (\ref{PsiTcompact}), but take the limit $Hb\gg1$ in the final step. The resulting expression is
		\beq
		dP\propto \frac{a^{2}}{b^{3}}\exp\left[-\frac{3\pi a^{2}}{2H^{2}b^{2}}\right]db,
		\label{Ptunnel}
		\eeq
		which is valid to lowest order in $(Hb)^{-1}$. Introducing a new variable $l$ which characterizes the relative size of $S^{1}$ and $S^{2}$, 
		\beq
		l=\frac{a}{b},
		\eeq
		we can express (\ref{Ptunnel}) as a distribution for $l^{2}$:
		\beq 
		dP\propto \exp\left[-\frac{3\pi l^{2}}{2H^{2}}\right]dl^{2}.
		\label{Pl}
		\eeq
		This is a normalizable distribution 
		with the average value
		\beq
		\Bar{l}^{2}=\frac{2H^2}{3\pi}.
		\eeq
		
		Thus, in the regime of large $a$ and $b$ the tunneling wave function predicts an ensemble of classical de Sitter-like universes (\ref{approxdS}) with $a/b\lesssim H\ll 1$.
		This is in contrast with the HH state which gives a distribution peaked at the Nariai solution (\ref{sol1}).

		\subsection{Small $S^{1}$ region $Ha\ll1$}
		
		In the KS model there is no clear distinction between classically allowed and forbidden regions, except when $Hb\approx1$. In this part of superspace the point $Ha=1$ divides the classical and quantum regimes. On the contrary, the classical universes (\ref{dS}) do not have a  well defined bounce point since $\dot{a}$ and $\dot{b}$ do not vanish simultaneously. Thus, we do not expect the wave function 
		to bear much resemblance to the usual tunneling picture established in quantum cosmology. 
		A related reason, mentioned in Sec.3, is that the WDW eq. is a hyperbolic equation, so the kinetic terms in the WDW operator have opposite signs. In this subsection we will try to obtain some insight into the behavior of the tunneling wave function close to the superspace boundary at $a=0$, without assuming $b$ to be small.

		\subsubsection{Saddles and contours}
		
		We are interested in finding the saddles in the region $u\ll1$. Setting $u=0$ in the action (\ref{RescaledSE}) and solving for the saddles through (\ref{saddleeq}) we obtain
		\beq
		N_{1,2}\approx-3\pm\sqrt{\frac{3v}{4-v}}
		\eeq
		where we have dropped the tildes. This approximation is valid as long as $N^{2}\gg3u$ and $v$ is not too close to 3. The first constraint justifies setting $u=0$ in the parenthesis of the third term in the action. The second constraint is imposed so that the first term in the action is larger than the second, which justifies dropping it for $u\ll1$. 
		
		The rest of the saddles can be approximately found by neglecting the third term in the action (\ref{RescaledSE}). Solving for the saddles yields
		\beq
		N_{3,4}\approx\pm\sqrt{\frac{3uv}{3-v}}.
		\eeq
		This approximation is valid everywhere in the region $u\ll v$, apart from values of $v$ that approach $v\approx3$, for which $N$ becomes large.
		
		Finally, in order to find the 5th saddle we will use the insight obtained from numerical results. The 5 saddles in the region $u\ll1$ and $v\sim1$ have the following characteristics. Two of them are ${\cal O}(1)$, the other two are ${\cal O}(\sqrt{u})$, while the 5th is ${\cal O}(u)$. In fact, we observe that when $v$ is not too small, the 5th saddle is given approximately by
		\beq
		N_{5}\approx-\frac{u}{2},
		\eeq
		which is in agreement with the lowest order expansion for $\tilde{N}_{1}$ given by Eq.(\ref{N12}). It is reassuring that all our saddle points can be matched in the appropriate limits in the different regions of KS superspace.
		
		Overall, we expect the above saddles to be valid when $u\ll1$, $u\ll v$, $v\neq\{3,4\}$. We must also note that a consequence of these restrictions is that $b$ is not allowed to approach zero.

		The analytic expressions for the saddles in the region $u\ll1$ suggest that there exist three qualitatively different steepest descent contour configurations depending on the value of $v$. In this subsection we are mostly interested in investigating the behavior of the wave function at small overall volume, so we will focus on the region $u\ll1$ and $v<3$. 
		
		In this case all five saddles are placed on the real-$N$ axis. The steepest descent contours are similar to the ones in Fig.\ref{fig4}. The only difference is that when $u\ll1$, the loop surrounding the singularity $N=0$ and the loop passing through $N=0$ are shrunk to a very small size. These loops are defined by the saddles $\mathcal{O}(\sqrt{u})$ and $\mathcal{O}(u)$ respectively, $N_{3,4}$ and $N_{5}$. The loop encircling the singularity $N=-3$ does not shrink, since it is defined by the saddles of zero order in $u$, $N_{1,2}$.
		
		The steepest descent/ascent lines for
		$u\ll1$ and $v<3$ are illustrated in Fig.\ref{fig4} . In this case the nearly Lorentzian contour can be distorted
		into a contour running from $N = 0$ along the upper arc to the saddle $N_{5}$, then along the
		real axis through the saddle $N_{4}$ until the saddle $N_{1}$. At that point it takes a turn and
		follows the upper arc to $N_{2}$, and runs from there to $N \rightarrow-\infty$. This contour is
		dominated by the saddle point $N_{4}$.
		
		\begin{figure}[h!]
			\centering
			\includegraphics[scale= 0.3]{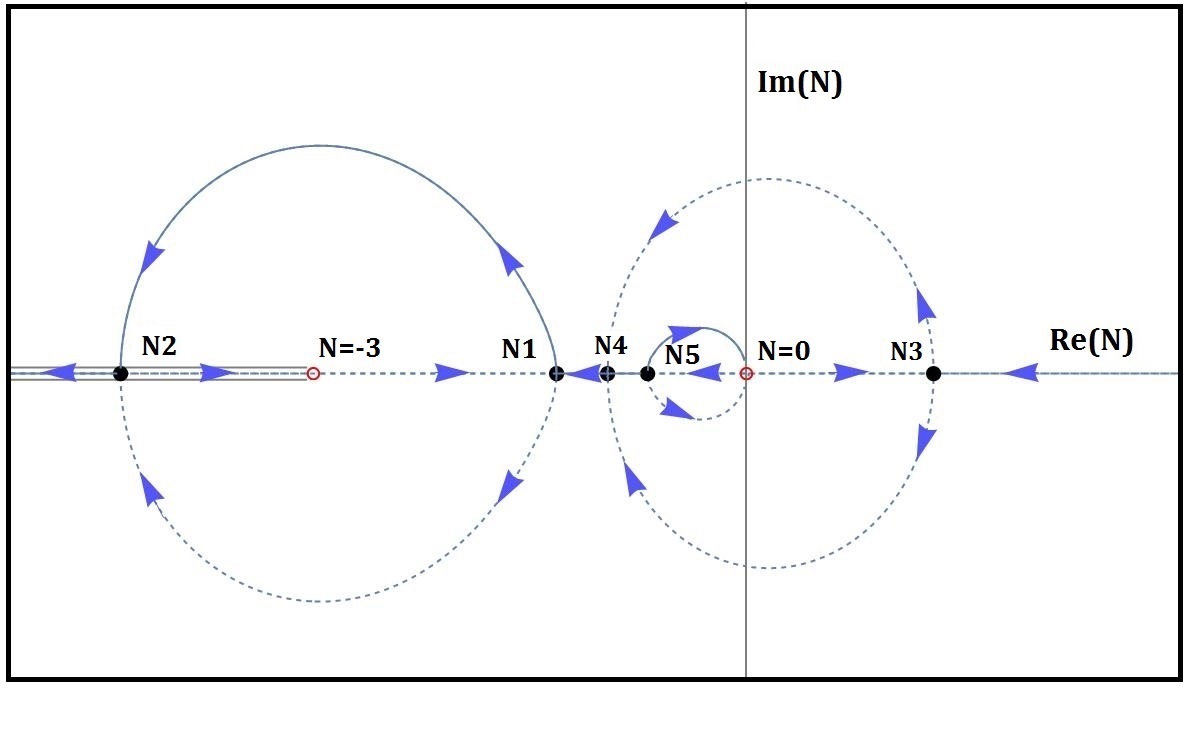}
			\caption{The steepest descent contours for $Ha\ll Hb<\sqrt{3}$. In this case, all the saddles are real.
				The tunneling contour corresponds to the solid curve starting from the singularity $N=0$ and dominated by saddle $N_{4}$.}
			\label{fig4}
			
		\end{figure}

		\subsubsection{WKB tunneling wave function}

		The WKB wave function corresponding to this saddle can be found along the same lines as previously demonstrated:
		\beq
		\Psi_{T}(Ha\ll1,Hb<\sqrt{3})\propto \exp\left(-2\pi ab\sqrt{1-\frac{H^{2}b^{2}}{3}}\right),
		\label{PsiT Ha<<1}
		\eeq
		where we have omitted the pre-exponential factor. This expression is valid in the region of superspace where $Ha\ll1$ and $Ha\ll Hb\neq\sqrt{3}$. We notice that for $a=$const,  $\Psi_T$ is a decreasing function of $Hb$ until it reaches a minimum at $Hb=\sqrt{3/2}$. For larger values of $Hb$ the wave function increases until it approaches $Hb\approx\sqrt{3}$ where our approximation breaks down. 
		
		In the case of $4>v>3$, the saddles $N_{1,2}$ become a complex conjugate pair, while the rest of the saddles remain real. Steepest descent analysis shows that the path integral receives dominant contributions from the saddles $N_{1}$ and $N_{5}$. Thus this region can be thought of as a transition from a fully quantum regime $v<3$ to a hybrid one, in which classical trajectories do penetrate, however they are accompanied by Euclidean components. 
		
		Finally, the case for which $u\ll1$, $v>4$ is qualitatively similar to the region $v\gg1$ discussed in section 5.3. So as long as $u\ll v$ the character of the steepest descent contours will not change and we are able to define a nearly-Lorentzian contour picking a contribution from the saddle $N_{1}$ as defined in Sec. 5.3 The WKB wavefunction will then be approximatelly given by the analytic continuation of (\ref{PsiT Ha<<1}) to $Hb>\sqrt{3}$ and will be valid for as long as $a\ll b$.
		
		Overall, the behavior of $\Psi_T$ that we have found is not in line with the familiar quantum tunneling through a barrier.  
		We note especially that the nucleation of a universe with $b\gg a$ is not exponentially suppressed.
		This can be seen from
		\beq
		\left|\frac{\Psi_{T}(Hb\gg Ha\gg1)}{\Psi_{T}(Ha\ll Hb<\sqrt{3})}\right|\sim1.
		\eeq
		On the other hand, exponential suppression is present in the narrow region $Hb\approx1$, where there is a clear bounce point at $Ha=1$ and 
		\beq
		\left|\frac{\Psi_{T}(Ha>1)}{\Psi_{T}(a\approx0)}\right|\approx\exp(-\frac{\pi}{H^{2}})\ll1.
		\eeq

		\subsection{Total nucleation probability}
		
		The total nucleation probability of an $S^{1}\times S^{2}$ universe can be found by integrating the distribution (\ref{Pl}) over $l^2$.  We obtain
		\beq
		P\propto H^2.
		\label{Pnuc}
		\eeq
		As in the de Sitter model, this probability is maximized at large values of $H$.  But the dependence on $H$ in (\ref{Pnuc}) is a power law, while in de Sitter model it is exponential \cite{Vilenkin:1987kf},
		\beq
		P_{dS}\sim \exp\left(-\frac{3\pi}{H^2}\right).
		\label{PdS}
		\eeq 
		
		If one adopts a minisuperspace framework that allows both $S^{1}\times S^{2}$ and $S^3$ topologies, then Eqs.(\ref{Pnuc}),(\ref{PdS}) suggest that for $H\ll 1$ nucleation of $S^{1}\times S^{2}$ universes is exponentially favored.  However, if different topologies are allowed, one can also consider adding a topological Gauss-Bonnet term to the action,
		\beq
		S_{GB}=\frac{\alpha}{16\pi}\int d^4 x\sqrt{-g}\left(R_{\mu\nu\sigma\tau}R^{\mu\nu\sigma\tau}-4R_{\mu\nu}R^{\mu\nu}+R^2\right) +S_{boundary}, 
		\label{GB}
		\eeq
		where $\alpha$ is a constant of dimension (length)$^2$ and $S_{boundary}$ is a boundary term which generalizes the Gibbons-Hawking term \cite{Myers}.  The addition of this term is necessary to make the boundary value problem well defined.  
		
		In $(3+1)$ dimensions the variation of the integrand in Eq.(\ref{GB}) is a total derivative, so the Gauss-Bonnet term has no effect on dynamics.  This term is a topological invariant,
		\beq
		S_{GB}=-2\pi\alpha\chi,
		\eeq
		where $\chi$ is the Euler character.  It can nevertheless have physical implications
		(see, e.g., \cite{Parikh} and references therein). In the context of quantum cosmology, the extra term (\ref{GB}) adds only a constant factor to the wave function, $f_{GB}=\exp({-2\pi\alpha\chi})$, if the topology of the universe is fixed.
		But if the topology is allowed to vary, different topologies will be weighted with different factors.
		The Euler character is $\chi=2$ for a de Sitter universe and $\chi=4$ for the $S^1\times S^2$ universe\footnote{These are the Euler characters for $S^4$ and $S^2\times S^2$ instantons, respectively.}.
		Hence the nucleation probabilities of such universes are
		\beq
		{P_{dS}}\sim \exp\left(-\frac{\pi}{H^2}-4\pi\alpha\right),
		\eeq
		\beq
		{P_{S^1\times S^2}}\sim H^2 \exp\left(-8\pi\alpha\right).
		\eeq
		This shows that de Sitter universes may dominate if $\alpha$ is sufficiently large,\footnote{A Gauss-Bonnet term (\ref{GB}) appears in the low-energy effective action of heterotic string theory \cite{Zwiebach}, together with an infinite series of higher-order curvature corrections
$\sim (\alpha R)^n$.  However, the higher-order terms can be neglected only if $\alpha R\ll 1$ \cite{Parikh}, which is in conflict with (\ref{Parikh}). Thus the Gauss-Bonnet term should have a different origin if it is to play a role in suppressing $S^1\times S^2$ universes.} 
\beq
\alpha >1/4H^2\gg 1.
\label{Parikh}
\eeq

		\section{Conclusions}
		
		In this project we applied the tunneling proposal for the wave function of the universe to the Kantowski-Sachs (KS) minisuperspace model of spatial topology $S^{1}\times S^{2}$.  
		The path integral version of this proposal defines the wave function $\Psi_T(g)$ as a path integral over histories interpolating between a vanishing 3-geometry ("nothing") and a given configuration $g$, with the lapse integration taken over semi-infinite Lorentzian contour.  
		It turns out, however, that all histories with vanishing initial radii of $S^1$ and $S^2$ in KS model necessarily have an initial curvature singularity, even after Euclidean continuation \cite{HL}.  It follows that the wave function defined in this way does not describe a non-singular origin of the universe, even at the semiclassical level.
		
		The conclusion could be that the tunneling wave function cannot be defined in the KS model.  This could mean that a universe of topology $S^{1}\times S^{2}$ cannot originate out of nothing -- assuming that the tunneling approach to the wave function of the universe is on the right track.
		
		Here we have explored an alternative possibility, introduced in Ref.\cite{HL} by Halliwell and Louko.
		They suggested that the boundary condition of vanishing 3-geometry should be replaced by the condition of a vanishing 3-volume.  That is, only one of the initial scale factors should be set equal to zero.
		The interpretation of such a degenerate but non-vanishing 3-geometry as "nothing" may be found objectionable.  On the other hand, such geometries can be obtained as limiting slices of regular Euclidean 4-geometries of topology $S^2\times S^2$, which may be regarded as instantons describing a non-singular origin of the universe.  We took an agnostic attitude to this issue and pursued the HL proposal in this paper, to see where it leads.   
		
		With one of the scale factors set to zero, the other one can be set at a nonzero value that is consistent with a non-singular 4-geometry.  Alternatively, leaving the other scale factor unspecified one can impose a regularity condition excluding conical singularities.  We have studied both of these approaches. 
		
		In the first approach, when the path integral is taken between fixed values of the scale factors with one of them set to zero, we found that the resulting wave function diverges at a finite value of the radius of $S^2$, $b=H^{-1}\sqrt{3}$.  Hence it is not a solution of the WDW equation in the entire superspace and is not a suitable choice for the tunneling wave function of the universe.  
		
		The main body of the paper is devoted to the second approach.  Here we found that the choice of $b=0$ cannot be supplemented by a regularity condition that would give an acceptable wave function.  One always gets the same singular wave function that was obtained for fixed initial scale factors.  The only option is then to set $a=0$ and impose a regularity condition of smooth closure on $S^1$.  We found that the resulting wave function is normalizable, with a probability distribution peaked at $b\gtrsim a/H \gg a$.  It predicts a highly anisotropic initial universe.  However, the universe expands exponentially in all directions, and after a large amount of inflation observers will see an isotropic local universe.  In contrast, the Hartle-Hawking wave function gives a probability distribution peaked at Nariai-type universes with $b\approx 1/H$, which remain locally anisotropic until late times.  
		
		We have emphasized that the wave function we obtained here is rather different from wave functions describing tunneling in quantum mechanics.  The reason is that the WDW equation of the KS model is a hyperbolic equation, with the two kinetic terms having opposite signs.  There is therefore no clear division of superspace into classically allowed and forbidden regions.  In particular, there is no exponential suppression of the probability distribution in the region of $b\gg a$.  
		
		We find that the total nucleation probability is $P\propto H^2$.  As in de Sitter model, it is maximized at large values of $H$, but unlike de Sitter the dependence on $H$ is a power law, not exponential.  
		It follows that for $H\ll 1$ the nucleation probability is much higher for $S^1\times S^2$ than for de Sitter universes.  We noted that this situation can be reversed if the gravitational action is supplemented with a Gauss-Bonnet term with a sufficiently large coefficient.
		
		It would be interesting to extend our analysis by including a homogeneous scalar field $\phi$, replacing the cosmological constant with a slowly varying potential $V(\phi)$.  
		Another interesting extension would be to study a "midi-superspace" model, including perturbatively an infinite number of inhomogeneous modes of a quantum field.  In the de Sitter model one finds that the universe nucleates with the field in a de Sitter invariant Bunch-Davies state, and in a more realistic model this initial state may lead to a nearly scale-invariant spectrum of density fluctuations, in agreement with observations.  On the other hand, in the KS model the initial quantum state will not have de Sitter, and not even rotational symmetry.  It is possible however that the field will approach the Bunch-Davies state at late times in the course of inflation, due to the no-hair "theorem" (e.g., \cite{Kaloper}).  
		
		We finally mention that KS model is closely related to the $2D$ Jackiw-Teitelboim gravity, which has recently attracted much attention.  This relation has been discussed in Ref.\cite{FV} in the case of the HH wave function, and it would be interesting to explore it for the tunneling wave function.

		\section*{Acknowledgements}
		
		We are grateful to Jorma Louko and Maulik Parikh for useful discussions.  This work was supported in part by the National Science Foundation under grant No. 2110466.


\section*{References}
		
		\begin{thebibliography}{}
			
			
			\bibitem{DeWitt:1967yk}
			B.~S.~DeWitt,
			``Quantum Theory of Gravity. 1. The Canonical Theory,''
			\href{doi:10.1103/PhysRev.160.1113}{Phys. Rev. \textbf{160}, 1113-1148 (1967)}
			
			
			
			\bibitem{Teitelboim}
			
			C. Teitelboim, "Quantum mechanics of the gravitational field". 
			\href{https://link.aps.org/doi/10.1103/PhysRevD.25.3159}{Phys. Rev. D \textbf{25}, 3159–3179 (1982)}
			
			\bibitem{Hartle:1983ai}
			J.~B.~Hartle and S.~W.~Hawking,
			``Wave Function of the Universe,''
			\href{https://link.aps.org/doi/10.1103/PhysRevD.28.2960}{Phys. Rev. D \textbf{28}, 2960-2975 (1983)}
			
			
			\bibitem{Vilenkin:1986cy}
			Vilenkin, A. (1986). Boundary conditions in quantum cosmology. \href{https://doi.org/10.1103/physrevd.33.3560}{Physical Review D, 33(12), 3560–3569. }
			
			
			
			
			\bibitem{Vilenkin:1987kf}
			A.~Vilenkin,
			``Quantum Cosmology and the Initial State of the Universe,''
			\href{ https://doi.org/10.1103/physrevd.37.888}{Phys. Rev. D \textbf{37}, 888 (1988)}
			
			\bibitem{Vilenkin:1994rn}
			A.~Vilenkin,
			``Approaches to quantum cosmology,''
			\href{https://doi.org/10.1103/physrevd.50.2581}{Phys. Rev. D \textbf{50}, 2581-2594 (1994)}
			
			\bibitem{Vilenkin:1982de}
			A.~Vilenkin,
			``Creation of Universes from Nothing,''
			\href{https://doi.org/10.1016/0370-2693(82)90866-8}{ Phys. Lett. B \textbf{117}, 25-28 (1982)}
			
			\bibitem{Linde:1983mx}
			A.~D.~Linde,
			``Quantum creation of an inflationary universe,''
			\href{https://doi.org/10.1007/bf02790571}{Sov. Phys. JETP \textbf{60}, 211-213 (1984)}
			
			\bibitem{Rubakov:1984bh}
			V.~A.~Rubakov,
			``Quantum Mechanics in the Tunneling Universe,''
			\href{https://doi.org/10.1016/0370-2693(84)90088-1}{Phys. Lett. B \textbf{148}, 280-286 (1984)}
			
			\bibitem{Vilenkin:1984wp}
			A.~Vilenkin,
			``Quantum Creation of Universes,''
			\href{https://doi.org/10.1103/physrevd.30.509}{Phys. Rev. D \textbf{30}, 509-511 (1984)}
			
			\bibitem{Zeldovich:1984vk}
			Y.~B.~Zeldovich and A.~A.~Starobinsky,
			``Quantum creation of a universe in a nontrivial topology,''
			Sov. Astron. Lett. \textbf{10}, 135 (1984)
			
			
			
			\bibitem{Halliwell:1988ik}
			J.~J.~Halliwell and J.~Louko,
			``Steepest Descent Contours in the Path Integral Approach to Quantum Cosmology. 1. The De Sitter Minisuperspace Model,''
			\href{doi:10.1103/PhysRevD.39.2206}{Phys. Rev. D \textbf{39}, 2206 (1989)}
			
			\bibitem{KS}
			
			R.~Kantowski and R.~K.~Sachs,
			``Some spatially homogeneous anisotropic relativistic cosmological models,''
			\href{https://doi.org/10.1063/1.1704952}{J. Math. Phys. \textbf{7}, 443 (1966)}
			
			
			\bibitem{FV}
			G.~Fanaras and A.~Vilenkin,
			``Jackiw-Teitelboim and Kantowski-Sachs quantum cosmology,''
			\href{doi:10.1088/1475-7516/2022/03/056}{JCAP \textbf{03}, no.03, 056 (2022)}
			
			
			
			
			\bibitem{Laflamme}
			R.~Laflamme, 
			"The wave function of a $S_1\times S_2$ universe", Ph. D. Thesis, University of Cambridge (1986).
			
			\bibitem{Louko:1988bk}
			J.~Louko,
			``Canonizing the Hartle-hawking Proposal,''
			\href{doi:10.1016/0370-2693(88)90008-1}{Phys. Lett. B \textbf{202}, 201-206 (1988)}
			
			\bibitem{Conradi}
			H.~D.~Conradi,
			``Quantum cosmology of Kantowski-Sachs like models,''
			\href{https://doi.org/10.1088/0264-9381/12/10/005}{Class. Quant. Grav. \textbf{12}, 2423-2440 (1995)}
			
			
			
			\bibitem{Louko:1988ia}
			J.~Louko and T.~Vachaspati,
			``On the Vilenkin Boundary Condition Proposal in Anisotropic Universes,''
			\href{doi:10.1016/0370-2693(89)90912-X}{Phys. Lett. B \textbf{223}, 21-25 (1989)}
			
			
			
			
			\bibitem{CH}
			G.~Conti and T.~Hertog,
			``Two wave functions and dS/CFT on S$^{1}$ \texttimes{} S$^{2}$,''
			\href{https://arxiv.org/abs/1412.3728}{JHEP \textbf{06}, 101 (2015)}
			
			
			
			\bibitem{HL}
			
			J.~J.~Halliwell and J.~Louko,
			``Steepest Descent Contours in the Path Integral Approach to Quantum Cosmology. 3. A General Method With Applications to Anisotropic Minisuperspace Models,''
			\href{https://doi.org/10.1103/physrevd.42.3997}{Phys. Rev. D \textbf{42}, 3997-4031 (1990)}
			
			
			
			
			\bibitem{Feldbrugge:2017kzv}
			J.~Feldbrugge, J.~L.~Lehners and N.~Turok,
			``Lorentzian Quantum Cosmology,''
			\href{https://arxiv.org/abs/1703.02076}{Phys. Rev. D \textbf{95}, no.10, 103508 (2017)}
			
			
			
			\bibitem{Bousso:1996au}
			R.~Bousso and S.~W.~Hawking,
			``Pair creation of black holes during inflation,''
			\href{doi:10.1103/PhysRevD.54.6312}{Phys. Rev. D \textbf{54}, 6312-6322 (1996)}
			
			
			
			\bibitem{Nariai}
			
			H. Nariai, ``On a New Cosmological Solution of Einstein’s Field Equations of Gravitation". \href{https://doi.org/10.1023/a:1026602724948}{General Relativity and Gravitation, \textbf{31(6)}, 963–971 (1999)}
			
			
			
			
			
			\bibitem{Vilenkin:1988yd}
			A.~Vilenkin,
			``The Interpretation of the Wave Function of the Universe,''
			\href{https://journals.aps.org/prd/abstract/10.1103/PhysRevD.39.1116}{Phys. Rev. D \textbf{39}, 1116 (1989)}
			
			\bibitem{Myers}
			R.~C.~Myers,
			``Higher Derivative Gravity, Surface Terms and String Theory,''
			Phys. Rev. D \textbf{36}, 392 (1987)
						
			\bibitem{Parikh}
			S.~Chatterjee and M.~Parikh,
			``The second law in four-dimensional Einstein-Gauss-Bonnet gravity,''
			\href{https://arxiv.org/abs/1312.1323}{Class. Quant. Grav. \textbf{31}, 155007 (2014)}.
			
			\bibitem{Zwiebach}
			B.~Zwiebach,
			``Curvature Squared Terms and String Theories,''
			Phys. Lett. B \textbf{156}, 315-317 (1985)

			
			\bibitem{Kaloper}
			N.~Kaloper and J.~Scargill,
			``Quantum Cosmic No-Hair Theorem and Inflation,''
			\href{https://arxiv.org/abs/1802.09554}{Phys. Rev. D \textbf{99}, no.10, 103514 (2019)}.
			
			
			
			
		\end{thebibliography}
	\end{document}